\documentclass[11pt]{article}

\usepackage[T1]{fontenc}

\usepackage{lineno,hyperref}
\modulolinenumbers[5]

\usepackage{amsmath}
\usepackage{amssymb}
\usepackage{amsthm}
\usepackage{tikz}
\usepackage{pgfplots}
\usepackage{graphicx}
\usepackage{slashbox}
\usepackage{caption}
\usepackage{subcaption}
\usepackage{xcolor}
\usepackage{balance}
\allowdisplaybreaks
\usepackage{algorithm}
\usepackage[noend]{algcompatible}
\usepackage{mathtools}
\usepackage{subcaption}
\usepackage{hyperref}
\usepackage{comment}
\usepackage{tabularx}
\usepackage{bbm}
\usepackage{thmtools}
\usepackage{cite}

\usepackage{hyperref}
\usepackage{pgfplots}
\usepackage{graphicx}
\usepackage{caption}
\usepackage{subcaption}
\usepackage{algorithm}
\usepackage[noend]{algcompatible}
\usepackage{multirow}
\usepackage{booktabs}
\usepackage{enumitem}
\usepackage{array}
\usepackage{authblk}
\usepackage{setspace}
\usepackage{fullpage,etoolbox}
\usepackage{color, colortbl}
\definecolor{Gray}{gray}{0.9}
\usepackage[most,breakable,many]{tcolorbox}
\tcbset{before skip=0pt,after skip=0pt} 

\allowdisplaybreaks

\renewcommand{\baselinestretch}{0.97}

\newcommand{\state}{s}

\newcommand{\gaussian}{\mathcal{N}}

\newcommand{\model}{M}

\newcommand{\horizon}{\mathrm{K}}
\newcommand{\timeid}{k}
\newcommand{\init}{\mathcal{I}}

\newcommand{\traj}{\sigma}

\newcommand{\relu}{\mathsf{ReLU}}

\newcommand{\zon}{\mathsf{Zonotope}}

\newcommand{\retosiduals}{\mathcal{R}_{D_{\text{test}}}^{j},\ j \in [n\horizon]}

\newcommand{\overallf}{\mathcal{F}}

\newcommand{\mypara}[1]{\vspace{0.5em} \noindent{\bf #1}.} 
\newcommand{\myipara}[1]{\vspace{0.5em} \noindent{\em #1}.} 

\def\relu{\mathrm{ReLU}}

\newcommand{\navid}[1]{#1}

\begin{document}
\newtheorem{theorem}{Theorem}
\newtheorem{definition}{Definition}
\newtheorem{lemma}{Lemma}

\title{Data-Driven Reachability Analysis of Stochastic Dynamical Systems with Conformal Inference }

\author[1]{Navid Hashemi}
\author[1]{Xin Qin}
\author[1]{Lars Lindemann}
\author[1]{Jyotirmoy V. Deshmukh}
\affil[1]{Thomas Lord Department of Computer Science, University of Southern California}

\maketitle

\begin{abstract}
We consider data-driven reachability analysis of
discrete-time stochastic dynamical systems using conformal inference. We assume that we are not
provided with a symbolic representation of the stochastic
system, but instead have access to a dataset of $\horizon$-step 
trajectories. The reachability problem is to construct a probabilistic flowpipe such that the probability that a $\horizon$-step trajectory can violate the bounds of the flowpipe does not exceed a user-specified failure probability threshold. The key ideas in this paper are: (1) to learn a surrogate predictor
model from data, (2) to perform reachability
analysis using the surrogate model, and (3) to quantify the surrogate model's incurred
error using conformal inference in order to give probabilistic reachability guarantees. We focus on learning-enabled control systems with complex closed-loop dynamics that are difficult to
model symbolically, but where state transition pairs can be queried, e.g., using a simulator. We demonstrate the applicability of our 
method on examples from the domain of 
learning-enabled cyber-physical systems.
\end{abstract}

\section{Introduction}


Reachability analysis of stochastic nonlinear dynamical systems is a challenging problem that has been 
extensively studied in the literature \cite{vinod2021stochastic,abate2008probabilistic,abate2007computational,
Maria,huang2019reachnn,huang2022polar,chen2013flow,dutta2019sherlock,koutsoukos2006computational,bansal2017hamilton,bortolussi2014statistical,kwiatkowska2007stochastic,legay2019statistical}. 
Most of the prior work is {\em model}-{\em based}, i.e., it requires a symbolic model of the
dynamical system which can then be over-approximated to obtain {\em flowpipes} or the set of reachable 
states of the system over a given time horizon. In this paper, we explore the notion of {\em model-free 
reachability analysis}, {\em i.e.}, to compute reachable sets of the stochastic dynamical system
even when we {\em do not have the symbolic system dynamics}, but have access to a numeric simulator or 
actual behaviors sampled from the system.  A significant 
advantage of such a data-driven technique is that we obtain not only (probabilistic) reachable sets for the
system or the simulation model from which the trajectories are sampled, but we can get results 
over the possibly {\em infinite} set of models/systems consistent with the set of sampled trajectories. 
This provides us the opportunity for an analysis technique that is robust to model uncertainty. 

There is growing literature on computing probabilistically approximate reachable sets directly from data.
The authors  in \cite{devonport2021data} utilize level sets of Christoffel functions and provide a technique to 
compute a high accuracy probabilistic reach-set for general nonlinear systems. On the other hand, \navid{as a comparison, }we only have access to sampled trajectories.
The authors in \cite{alanwar2023data} propose specific parametric models (linear or polynomial),
to identify the Markovian stochastic dynamics of the system from data, and then perform reachability analysis on
the identified models. In contrast, we learn non-parameteric surrogates (using neural networks), which are not
restricted to Markovian dynamics, and quantify uncertainty using conformal inference.
In \cite{devonport2020data}, the authors use an interesting approach based on a Gaussian process-based
classifier to separate reachable states from unreachable states, and approximate the reach set by 
computing the sublevel set of the classifier. We note that this approach uses adaptive sampling of 
initial states where states are chosen to reduce uncertainty of the surrogate. This may require solving
a high-dimensional optimization problem and does not give probability guarantees. The authors also
propose an interval abstraction of the reach set, where sample complexity bounds are provided; however,
this approach may suffer from conservatism and high computational cost in high-dimensional systems.
In \cite{fisac2018general}, the authors assume partial knowledge of the model, while using data to deal
with Lipschitz-continuous state dependent uncertainty. 
The authors in \cite{lin2023generating} propose a method called DeepReach that uses a neural PDE solver 
to perform Hamilton-Jacobi method-based reachability analysis for high-dimensional systems. Here,
the complexity of the flowpipe computation scales with the complexity of the flowpipe itself
rather than the system dimension. While this method uses neural methods to perform reachability
analysis, it still requires access to the system dynamics. In \cite{dryvr}, the authors combine
simulation-guided reachability analysis techniques with data-driven techniques. Here, the authors
estimate a discrepancy function using system trajectories using a probably-approximately-correct 
learning algorithm. The discrepancy function bounds the distance between system trajectories
as a function of the distance between their initial states. This function is then used to inflate
the simulation trajectories to obtain reachtubes that are then used to compute the approximate
reachable set of states. We remark that such an approach requires a parametric form of the discrepancy
function which could be hard to obtain.

Our approach to compute probabilistic reachable sets for \navid{stochastic} systems has 
the following main steps: (1) we sample a number of system trajectories according to the 
user-provided distribution on the set of initial states, (2) we learn a data-driven
surrogate model that predicts the next $\horizon$ states of the system from a given state, 
(3) we perform traditional set propagation-based reachability analysis on the surrogate model,
and (4) we inflate the resulting reachable sets in a systematic fashion to account for the 
uncertainty induced by sampling from the user-provided distribution on the set of initial states and system stochasticity. 

While our general technique can work with arbitrary regression-based surrogate models,
in this paper, we choose feedforward neural networks as a data-driven model of the system
dynamics\footnote{There are many techniques to identify probably approximately correct
surrogate models, for example, using concentration bounds for system identification
\cite{matni2019tutorial}. In this paper, we eschew such methods, instead preferring the
data-efficient approach for uncertainty quantification provided by conformal inference.}. A key reason for choosing feedforward neural networks is 
to use recently developed techniques to perform set-propagation 
through neural networks \cite{tran2020nnv}. However,
since the surrogate model is learned by sampling, we still need to account for this uncertainty. 
To do this, we rely on {\em conformal
inference} (CI) or {\em conformal prediction}~\cite{vovk2005algorithmic,lei2014distribution}. In the machine learning community, CI is a well-known data-efficient framework for error analysis 
of predictive models. Recently, CI has been employed for data-driven stochastic verification  \cite{bortolussi2019neural,cairoli2023conformal,lindemann2023conformal,qin2022statistical}. For 
instance, in \cite{cairoli2023conformal,qin2022statistical,lindemann2023conformal}, the authors obtain 
confidence intervals on the robustness degree of system behaviors (for example based on the quantitative 
satisfaction of a given signal temporal logic (STL) property). A na\"{i}ve adaptation of existing
techniques for STL robustness verification to reachability analysis would require training surrogate models 
to predict each state variable independently (effectively discarding the rich correlation between
the state variables) resulting in an analysis that is too conservative. Instead, our approach 
preserves the structure of the dynamics through our usage of a surrogate model, while still 
accounting for uncertainty in a systematic fashion using CI.

The layout of our paper is as follows.  In Section \ref{sec:prelim}, we present the preliminaries 
and problem statement. Our algorithm for probabilistic reachable set computation is proposed in 
Section~\ref{sec:verification}. We demonstrate the efficacy of our method on several numerical 
examples in Section~\ref{sec:results}, and we finally conclude in Section~\ref{sec:conclusion}.

\mypara{Notation} We use bold letters to indicate vectors and vector-valued functions, and 
calligraphic letters to denote sets. We denote the set $\left\{1,2,\cdots, n \right\}$ with $[n]$. 
We present the structure of a feedforward neural network (FFNN) with $\ell$  hidden layers as 
an array $[n_0,n_1,\cdots n_{\ell+1}]$, where $n_0$ denotes the number of inputs $n_{\ell+1}$ 
is the number of outputs and $n_i , i\in [\ell]$ denotes the width of the $i$-th hidden layer. 
We denote $e_i\in\mathbb{R}^n$ as the $i$-th base vector of $\mathbb{R}^n$. The expression $x \sim 
\mathcal{X}$ means, the random variable $x$ follows the distribution $\mathcal{X}$. We denote the
cardinality of a set $A$ with $|A|$. $\zon(b, A)$ denotes a zonotope~\cite{kurzhanski2002ellipsoidal} 
with center $b$ and array of base vectors $A$. The operator $\oplus$ denotes the Minkowski sum. We
also denote $\lceil x \rceil$ as the smallest integer greater than $x \in \mathbb{R}$.



\section{Problem statement and Preliminaries}
\label{sec:prelim}

\noindent \textbf{Stochastic Dynamical System.} Estimating the set of reachable sets of
stochastic dynamical systems starting from a compact set of initial states $\init \subset \mathbb{R}^n$
is a well-studied problem. While this problem has been studied largely with model-based techniques,
we focus on {\em bounded-time, model-free reachability} analysis of a black-box discrete-time
stochastic dynamical system $M$. A random trajectory of $M$ can be modelled as a sequence of 
time-stamped states $\traj_{\state_0} = [\state_0^\top, \cdots, \state^\top_\horizon]^\top \subset 
\mathbb{R}^{(\horizon+1)n}$. We denote the distribution over the set of trajectories consistent
with $M$ as $\mathcal{S}$, and use $\traj_{\state_0} \sim \mathcal{S}$ to denote that $\traj_{\state_0}$
is sampled from $\mathcal{S}$. The distribution $\mathcal{S}$ could be induced due to a distribution
on the set of initial states of the system as well as stochastic uncertainties in its dynamics. 
For example, the transition dynamics could be Markovian, i.e.,
$\state_{\timeid} \sim T(\state' \mid \state = \state_{\timeid-1})$. However, we remark
that the techniques proposed in this paper can work for systems with non-Markovian dynamics.
For convenience, we denote the distribution over the set of initial states $\init$ as $\mathcal{W}$, and
assume that $(\Pr[\state_0 \notin \init]=0)$. The notation $\state_0 \stackrel{\mathcal{W}}{\sim} \init$
is used to denote that the state $\state_0$ is sampled from $\init$ using the distribution $\mathcal{W}$.
A practical choice of $\mathcal{W}$ may be to uniformly sample $\init$.  

\textbf{Trajectory Datasets. } We assume that we have access to a dataset of 
independently sampled trajectories of $M$, where the initial states are sampled with 
$\state_0 \stackrel{\mathcal{W}}{\sim} \init$. We assume that each trajectory has 
$\horizon$ time-steps. We split the trajectory dataset into two separate datasets, 
denoted as $D_{\text{train}}$ and $D_{\text{test}}$. 
Here $D_{\text{train}}$ is our training dataset and is to be used for training a surrogate model of $M$. On the other 
hand, $D_{\text{test}}$ is the test dataset, and is to be utilized  to generate the calibration dataset for deriving 
statistical properties. Let $D_{\text{test}}=\{\state_{0,i}, \traj_{\state_0,i}\}_{i=1}^L$ where $L$ is the number 
of test trajectories. 
 Our goal is now to compute a probabilistic reach set such that 
 $\traj_{\state_0}$ is included inside with a probability not lower than a user-defined
 threshold $1-\varepsilon$, where $\varepsilon \in (0,1)$.

We remark that although the data-points within a single trajectory are not independent and identically distributed (i.i.d.), we observe that the trajectory $\traj_{\state_0}$ can be seen as a vector $\traj_{\state_0} \in \mathbb{R}^{n(\horizon+1)}$ so that entire trajectories within $D_{\text{train}}$ and $D_{\text{test}}$ 
can be viewed as i.i.d. samples in the $\mathbb{R}^{n(\horizon+1)}$-space.
This observation is crucially used later when we quantify uncertainty using conformal prediction. 

 \textbf{Surrogate Model. } After obtaining the trajectory dataset, we train a surrogate model $\overallf: \mathbb{R}^n \to \mathbb{R}^{(\horizon+1) n}$ that maps a given initial state to the predicted $\horizon$-step
trajectory of the system. For instance, the surrogate model can be a feedforward neural network with $n$ inputs
and $(\horizon+1)n$ outputs. Let $\bar{\traj}_{\state_0}$ denote the predicted trajectory, then, \(
 \bar{\traj}_{\state_0} = \overallf(\state_0).\)
 Training such a $\horizon$-step predictive model can be difficult, especially when the dynamics are 
 nonlinear. Hence, in practice, we train models that take as input trajectory fragments to predict
 future trajectory fragments and then sequentially compose such models to obtain a longer predicted 
 trajectory. We remark that this does not affect the probabilistic reasoning that we perform later
 in the paper as the entire predictive model can still be treated as a deterministic map that
 takes as input the initial state $\state_0$ and outputs a $\horizon$-step predicted trajectory.

 \textbf{Conformal Inference.} Conformal inference \cite{vovk2005algorithmic,lei2014distribution,lei2018distribution} is a statistical tool for uncertainty quantification that has recently been used for analysing the uncertainty in the predictions performed by complex machine learning models \cite{angelopoulos2021gentle,lindemann2023safe,luo2022sample}. Conformal inference/prediction performs error quantification without making any assumption on the underlying data-generating distribution or the machine learning model. Consider a regression model $\mu$ and the random variables $z_1,z_2,...,z_{m+1}$ where $z_i=(x_i,y_i) \in \mathbb{R}^n \times \mathbb{R}$ with $ i\in [m+1]$ which are all independently sampled from the same distribution. Given a miscoverage level $\alpha \in (0, 1)$, conformal inference enables us to compute a prediction interval $C(x_{m+1})=[\mu(x_{m+1})-R^*,\ \mu(x_{m+1})+R^*] \subset \mathbb{R}$ from $z_1,z_2,...,z_{m}$ such that,
$$
\Pr[ y_{m+1} \in C(x_{m+1})] \geq 1- \alpha.
$$
More formally, define the residual  $R_i=\mid y_i-\mu(x_i) \mid$ for all $z_i$ with $i\in [m+1]$. Since the random variables $z_1,z_2,...,z_{m+1}$ are independent and identically distributed, the same applies to the residual $R_1,\hdots, R_{m+1}$. Under the assumption that $m$ satisfies $\ell=\lceil (m+1)(1-\alpha)\rceil\le m$, it holds that $R^*$ can be chosen to be the $\ell$-th smallest residual \cite[Lemma 1]{tibshirani2019conformal}. Without loss of generality, if $R_1,\hdots, R_{m}$ are sorted in non-decreasing order, then $R^*=R_\ell$ and it holds that $\Pr[ R_{m+1} \leq R^*] \geq (1-\alpha)$. By the choice of the residual, it hence holds that
$$
\Pr\Big[ y_{m+1} \in [\mu(x_{m+1})-R^*,\ \mu(x_{m+1})+R^*] \Big] \geq 1-\alpha.
$$
We also refer to $\delta = 1-\alpha$ as the confidence probability.

 \textbf{Problem Definition.} Our probabilistic reachability analysis can be formally stated as follows. Given the stochastic dynamical system $M$ with initial state $\state_0 \stackrel{\mathcal{W}}{\sim} \init$ and trajectory distribution $\mathcal{S}$, training and test datasets $D_{\text{train}}$ and $D_{\text{test}}$ consisting  of $\horizon$-step trajectories independently sampled from $\model$,  and a  user provided failure probability threshold $\varepsilon \in (0,1)$, compute a probabilistic reach set (also called flowpipe) $X$ such that:
\begin{equation}\label{eq:verific_prob}
 \Pr\left[ \traj_{\state_0} \in X \right]  \geq  1-\varepsilon
\end{equation}


\section{Scalable Data-Driven Reachability}
\label{sec:verification}

In the setting described in the previous section, we now show how to compute a reach set or a flowpipe $X\subset \mathbb{R}^{n(\horizon+1)}$ using reachability analysis and conformal inference. This flowpipe will contain the trajectory $\traj_{\state_0}$ of $M$ sampled with initial state $\state_0 \stackrel{\mathcal{W}}{\sim} \init$ with the confidence level of $\Delta = 1-\varepsilon$. We hence denote this flowpipe as $\Delta$-confident flowpipe and  define it as follows. 

\begin{definition}[$\Delta$-confident flowpipe] For a given confidence probability $\Delta\in (0,\ 1)$ and  a random trajectory $\traj_{\state_0} \sim \mathcal{S}$ with initial state $\state_{0} \stackrel{\mathcal{W}}{\sim} \init$, we say that $X \subset \mathbb{R}^{n(\horizon+1)}$ is a $\Delta$-confident flowpipe if
\begin{equation}\label{eq:confreg}
\Pr[\traj_{\state_0} \in X ] \geq \Delta
\end{equation}
    
\end{definition}
In this work, we are interested in computing $X$ while our access to $M$ is limited to the datasets $D_{\text{train}}$ and $D_{\text{test}}$. We will show that we can compute $X$ with valid guarantees by employing reachability analysis on the surrogate model trained from $D_{\text{train}}$ and error analysis of this model by applying conformal prediction on $D_{\text{test}}$. 

\subsection{Computing Reachsets for Surrogate Models}
\subsubsection{ReLU Surrogate Model}
We start with training a  neural network surrogate model $\overallf: \mathbb{R}^n \to \mathbb{R}^{n(\horizon+1)}$ over the training dataset $D_{\text{train}}$. The surrogate model is trained to approximate the trajectory $\traj_{\state_{0}} \in \mathbb{R}^{n(\horizon+1)}$ of the stochastic system $M$ sampled with initial state $\state_0 \stackrel{\mathcal{W}}{\sim} \init$. We denote the prediction $\bar{\traj}_{\state_{0}} \in \mathbb{R}^{n(\horizon+1)}$ of $\traj_{\state_{0}}$ as,  
$$
\bar{\traj}_{\state_{0}} = \overallf(\state_{0}) = 
\left[ \state_0^\top ,\ \mathsf{F}^1(\state_0) , \  \cdots ,\ \mathsf{F}^n(\state_0), \  \cdots , \  \mathsf{F}^{(n-1)\horizon}(\state_0) ,\  \cdots ,\  \mathsf{F}^{n\horizon}(\state_0)\right]^\top
$$
where $\mathsf{F}^j(\state_0)$ is the $(j+n)$-th component of the vector $\overallf(\state_0)$. 

Recent works in the literature have had great success on obtaining accurate bounds for the reachability analysis of $\relu$ neural networks  using polyhedral sets  \cite{tran2019star,tran2019safety,tran2020nnv}. The accuracy of these techniques motivates us to use $\relu$ activation functions for training neural networks as the surrogate models. These surrogate models will be used for deterministic reachability analysis which provides surrogate flowpipes which we formally define next. 

\begin{definition}[Surrogate flowpipe] 
The surrogate flowpipe $\bar{X}\subset \mathbb{R}^{n(\horizon+1)}$ contains the image of $\overallf(\init)$. Formally, for all, $\state_0\in \init$ it has to hold that
$\overallf(\state_0) \in \bar{X}$.   
\end{definition}

The reachability analysis methodology for $\relu$ neural networks in \cite{tran2020nnv} introduces two different approaches known as the exact-star and approx-star techniques. These are used to compute a surrogate flowpipe $\bar{X}$. The exact-star technique proposes exact reachability analysis using star sets, but can be slower due to its inherent computational complexity. On the other hand, the approx-star technique computes over-approximation of the flowpipe and is thus runtime-efficient, although it may make the surrogate model reachable set estimation conservative. The computational complexity of the exact-star technique and the conservatism of the approx-star technique can both be noticeably reduced using the idea of set partitioning \cite{tran2020nnv}. In this approach, we partition the set of initial conditions $\init$ into $N$ different sub-partitions,
$$
\init_i \subset \init , \quad \bigcup_{i=1}^N \init_i =\init, 
$$
and perform reachability analysis on every single sub-region with parallel computing. The inclusion of set-partitioning results in noticeable improvement in the computational efficiency and helps us compute more
accurate $\Delta$-confident flowpipes.


\subsection{Computation of a guaranteed $\Delta$-confident flowpipe}

When we train neural network surrogate models, typically we minimize a loss function defined as the difference between the $\horizon$-step trajectory predicted by the surrogate model and the actual trajectory. Depending on the dynamics of the underlying stochastic system $M$ and depending on how well the surrogate model is trained,
there is potential for error when predicting the trajectory from a previously unseen initial state. To give probabilistic bounds on this error, we 
utilize conformal prediction. We formally define the notion of the residual error as follows. 
\begin{definition}[Residual Error]\label{def:residual} 
Given a realization $(\state_0,\ \traj_{\state_0})$  sampled from the stochastic black-box system $\model$, the residual $R^j \in \mathbb{R}_{>0}$ is the distance between the $(j+n)$-th component of the trajectory and $\mathsf{F}^j(\state_0)$\footnote{Note that we denote the trajectory as a single row vector where component-wise states at each time-stamp are concatenated. Thus, for the $i^{th}$ state variable at time $\timeid$, $j$ will
be equal to $i\cdot \timeid + n$ (skipping the first $n$ component-wise values corresponding to the initial state $\state_0$).} Formally, we define
\[
R^j = \left|e_{j+n}^\top \traj_{\state_0} - \mathsf{F}^j(\state_0)\right|
\]
where, $e_j \in \mathbb{R}^{n(\horizon+1)}$ is the $j^{th}$ base vector of $\mathbb{R}^{n(\horizon+1)}$.
\end{definition}
We consider the component-wise residual $R^j$ since every single component $e_{j+n}^\top \traj_{\state_0}$ in $\traj_{\state_0}$ may represent a different quantity at a different time. For example, it would not make sense to define a joint error of the position and velocity of a system at some time, which motivates the definition of the component-wise error. 
Now that we have defined the residual $R^j$ for a component $j$, let us compute this residual for all calibration trajectories from $D_{\text{test}}$, i.e., for $\traj_{\state_{0,i}}, i \in [L]$.

\begin{definition}[Calibration Dataset] 
For a given test dataset $D_{\text{test}}$, the calibration dataset $\retosiduals$, is a collection of pairs $(\state_{0,i}, R^j_{i}), i\in[L]$, such that
$$
\begin{aligned}
\mathcal{R}_{D_{\text{test}}}^{j} = \Big\{ \left( \state_{0,i}, R^j_i \right) \mid \ &  (\state_{0,i}, \traj_{\state_{0,i}}) \in D_{\text{test}}, \; i\in [L] \Big\}.
\end{aligned}
$$
where $R^j_i = \mid e_{j+n}^\top \traj_{\state_{0,i}} - \mathsf{F}^j(\state_{0,i})\mid $.
\end{definition}



As our access to the underlying system $M$ is limited to a finite number of pre-recorded trajectories, it is not possible to exactly compute the distribution of the residual $R^j, j \in [n\horizon]$. Consequently, we cannot   compute the $\delta$-quantile of $R^j, j \in [n\horizon]$.\footnote{The $\delta$-quantile of a random variable $R^j$ is defined as $\inf\{z\in\mathbb{R} | \text{Pr}[R^j\le z]\ge  \delta\}$ for $\delta \in (0,1)$.}  However, we can utilize the method of conformal prediction \cite{vovk2005algorithmic} to compute  an upper bound on the $\delta$-quantile of $R^j$. Based on the technique introduced in Section~\ref{sec:prelim}, we sort the  residuals $R^j_i$ in non-decreasing order. For simplicity and without loss of generality, let us re-index the values of $R^j_1,\hdots, R^j_L$ so that $R^j_1 \leq R^j_2 \leq \cdots \leq R^j_L$. We can now apply conformal prediction and define $R^{j*} = R^j_\ell$ where $\ell = \lceil (L+1 ) \delta \rceil$. Consequently, we know that
$$
\Pr[R^j \leq R^{j*}] \geq \delta.
$$
In other words, the $\ell$-th smallest residual $R^{j*}$ is an upper bound for the $\delta$-quantile of the random variable $R^j$.



Recall that our ultimate goal is to compute a $\Delta$-confident flowpipe  $X$ for trajectories $\traj_{\state_0}$ from $M$ with distribution $\state_0 \stackrel{\mathcal{W}}{\sim} \init$. To that end, we compute the vector of upper bounds for all components of residual's $\delta$-quantile from conformal inference, and we denote it by $R^* = \left[ R^{1*},\cdots,R^{{n\horizon}*} \right]^\top$.



\begin{theorem}\label{lem:inclusion_conformal}
Let $\bar{X}$ be a surrogate flowpipe of the surrogate model $\overallf$ for the set of initial conditions $\init$. Let $R^{j*}$ be computed from the calibration dataset $\retosiduals$ with confidence probability $\delta\in (0,1)$ where $\mathcal{R}_{D_\text{test}}^j$ is based on i.i.d. test trajectories in $D_{\text{test}}$ from the stochastic system $M$ with initial state $\state_0 \stackrel{\mathcal{W}}{\sim} \init$.  Define the inflated surrogate flowpipe,
$$
X = \bar{X} \oplus \zon(0, \mathbf{diag}([0_{1\times n},\ R^*])),\ 
R^* = \left[ R^{1*},\cdots,R^{{n\horizon}*} \right].
$$
Then, it holds that $X$ is a $\Delta$-confident flowpipe with $\Delta = 1-n\horizon(1-\delta)$ for $\traj_{\state_0}$ from $M$ with initial state $\state_0 \stackrel{\mathcal{W}}{\sim} \init$.

\end{theorem}

\begin{proof}
Based on the definition of the conformity score, we have $\Pr\left[ R^j \leq R^{j*} \right] \geq \delta$. Therefore, we have that
$$
\Pr[R^j > R^{j*}] < 1-\delta.
$$
By applying the union bound over probabilities, it follows that
$$
\Pr\left[  \bigvee_{j=1}^{n\horizon} \left(R^j > R^{j*} \right) \right] <  n\horizon(1-\delta).
$$
The negation of this statement implies that
\begin{equation}\label{eq:0}
\Pr\left[  \bigwedge_{j=1}^{n\horizon} \left(R^j \leq R^{j*} \right) \right] \geq  1-n\horizon(1-\delta).
\end{equation}
We now denote $\Delta= 1-n\horizon(1-\delta)$ so that we can rephrase the above statement as, 
$$
\Pr\left[  \bigwedge_{j=1}^{n\horizon} \left( \mid e_{j+n}^\top \traj_{\state_0} - \mathsf{F}^j(\state_0) \mid \leq  R^{j*}\right) \right] \geq  \Delta.
$$

Next, we define the interval $C_j(\state_0)$ = $\left[ \mathsf{F}^j(\state_0)  -R^{j*}\ ,\  \mathsf{F}^j(\state_0)  +  R^{j*} \right]$. Accordingly,  we have
$$
\Pr\left[  \bigwedge_{j=1}^{n\horizon} \left( e_{j+n}^\top \traj_{\state_0} \in C_j(\state_0) \right) \right]  \geq  \Delta
$$
Based on this, we can now see that
\begin{equation} \label{eq:1}
\Pr\left[  \traj_{\state_0} \in \zon\left(\overallf(\state_0) , \mathbf{diag}\left([0_{1\times n},\ R^*]\right)\right) \right]\geq \Delta
\end{equation}
Since $\state_0 \stackrel{\mathcal{W}}{\sim} \init$ and $\bar{X}$ is a surrogate flowpipe for the surrogate model $\overallf$ on $\init$, i.e., $\state_0 \in \init$ implies $\overallf(s_0) \in \bar{X}$, we can conclude,
\begin{equation}\label{eq:2}
\zon\left(\overallf(\state_0) , \mathbf{diag}\left([0_{1\times n},\ R^*]\right)\right)\ 
 \subset \bar{X} \oplus \zon\left(0 , \mathbf{diag}\left([0_{1\times n},\ R^*]\right)\right) = X
\end{equation}
This fact implies that $\Pr[\traj_{\state_0} \in X  ] \geq \Delta$, i.e., $X$ is a $\Delta$-confident flowpipe, which completes the proof.
\end{proof}

Theorem \ref{lem:inclusion_conformal} tells us how to obtain a  $\Delta$-confident flowpipe given the $\horizon$-step datasets $D_{\text{train}}$ and $D_{\text{test}}$. We can now compute a lower bound on the minimum size of the calibration dataset that we need given a confidence probability $\Delta\in (0,1)$. Specifically, we note that $\Delta= 1-n\horizon(1-\delta)$  is equivalent to $\delta= 1- \frac{1-\Delta}{n\horizon}$.  The minimum required size $L$ of the calibration dataset has to satisfy $\lceil (L+1)\delta \rceil \leq L $ which gives us the explicit lower bound $L \geq \lceil \frac{1+\delta}{1-\delta} \rceil$. In this work, we defined the residual component-wise, recall Definition \ref{def:residual}. As observed in the proof of Theorem \ref{lem:inclusion_conformal}, we thus had to apply the union bound over all residuals $R^j$. This may in some cases be conservative, i.e., for large system dimension $n$ or large trajectory horizon $\horizon$. However, there are possible ways to define a residual in a way that removes this conservatism. For example, in our recent work \cite{cleaveland2023conformal} we show how to obtain tight conformal prediction regions for time series. Applying this method will also result in better data efficiency. We intend to explore this method in the context of this paper in future work and refer the reader to \cite{cleaveland2023conformal} for more details.



\section{Experimental Results}
\label{sec:results}
\begin{figure*}
    {\includegraphics[width=1.0\linewidth]{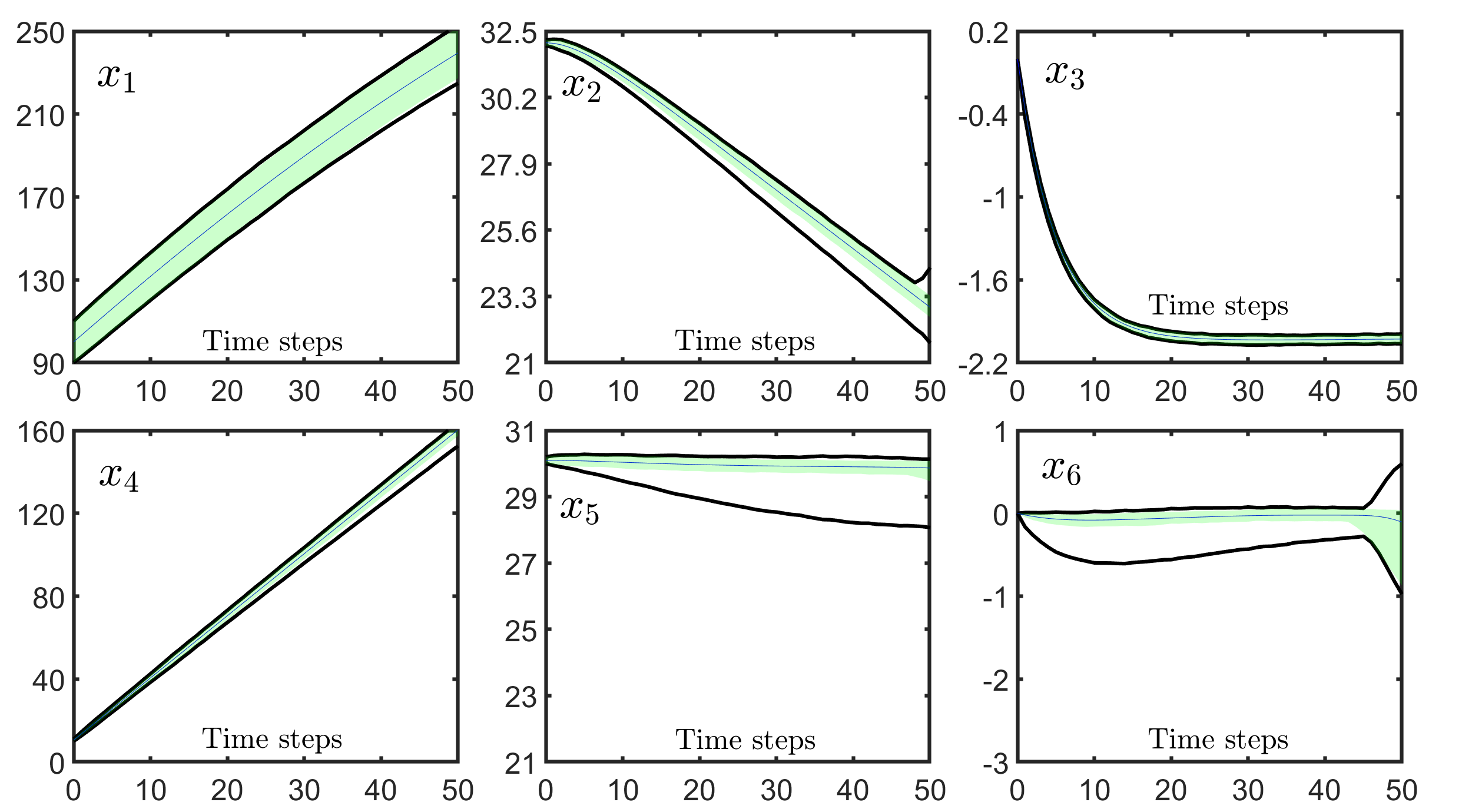}}
    {\caption{Adaptive Cruise Control: The black lines indicate the probabilistic flowpipe computed using the exact-star technique on the learned surrogate model combined with conformal inference with failure probability $\varepsilon=0.01$. The green shaded areas and blue lines are computed from $100000$ random trajectories from the ODE model. The green area shows the maximum and minimum value of the trajectory components over this dataset, and the blue line represents their average value.}
    \label{fig:acc_comparison}}
\end{figure*}

\begin{table}[t]
    \resizebox{\textwidth}{!}{
    \begin{tabular}{ccc||ccc} 
    \toprule
    Failure & Conformal inference & Reachability& Failure  & Conformal inference & Reachability  \\
     Probability & Run-time & Run-time & Probability & Run-time & Run-time \\ 
    \midrule \rowcolor{Gray}
    $0.10$ &   2.9439 $\mathrm{sec}$ & 9.8059 $\mathrm{sec}$ &  $0.05$ & 2.8783 $\mathrm{sec}$ &10.7457 $\mathrm{sec}$  \\
    $0.09$ &  2.8511 $\mathrm{sec}$ &9.0679 $\mathrm{sec}$  &  $0.04$ & 2.7901 $\mathrm{sec}$ &10.0033 $\mathrm{sec}$  \\
    \rowcolor{Gray}
    $0.08$ &  2.7367 $\mathrm{sec}$ &10.0502 $\mathrm{sec}$  &  $0.03$ & 3.096 $\mathrm{sec}$ &9.4364 $\mathrm{sec}$  \\
    $0.07$ &  2.8785 $\mathrm{sec}$ &9.7935 $\mathrm{sec}$ &  $0.02$ & 3.3982 $\mathrm{sec}$ &10.0027 $\mathrm{sec}$  \\
    \rowcolor{Gray}
    $0.06$ &  2.9341 $\mathrm{sec}$ &9.7988 $\mathrm{sec}$  & $0.01$ & 2.8536 $\mathrm{sec}$ &9.7945 $\mathrm{sec}$  \\
    \bottomrule
    \end{tabular}
    }
    \caption{Adaptive Cruise Control: Computation times of our method for different user-provided failure probabilities $\varepsilon$. The reachability run-time is the overall time for reachability analysis over the surrogate model and constructing the probabilistic flowpipe with conformal inference. The run-time for training a surrogate $\relu$ model was 2 hours and the run-time for test data-generation was 122 seconds.}  \label{tbl:acc}
\end{table}

\begin{figure*}[t]
    \hspace{-8mm}
    {\includegraphics[width = 1.1\textwidth]{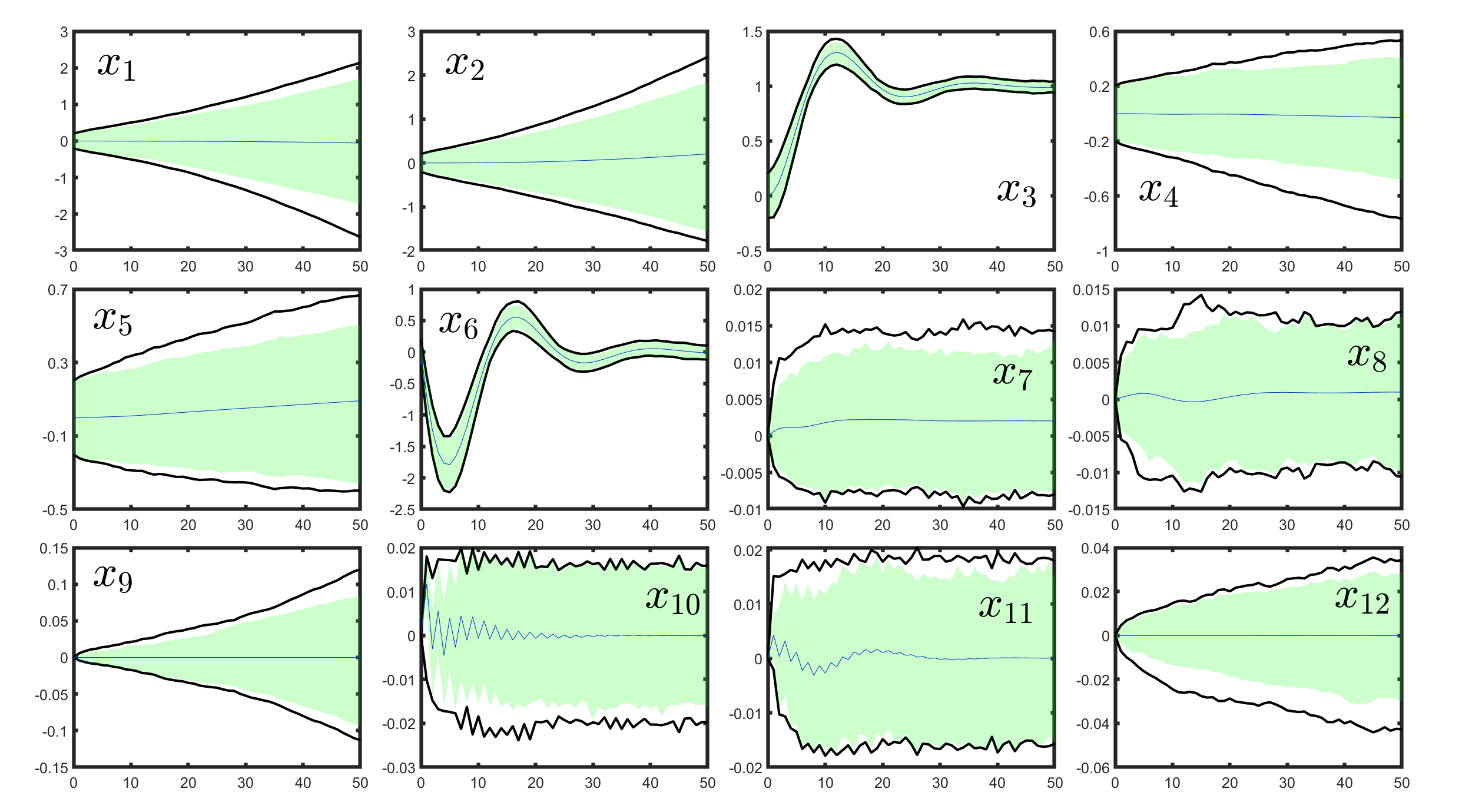}}
    {\caption{Quadcopter: The black lines indicate the probabilistic flowpipe computed using the exact-star technique on the learned surrogate model combined with conformal inference with failure probability $\varepsilon=0.01$. The green shaded areas and blue lines are computed from $100000$ random trajectories from the ODE model. The green area shows the maximum and minimum value of trajectory components over this dataset, and the blue line represents their average value.}
    \label{fig:Quad}}
\end{figure*}

We consider an adaptive cruise controller, a quadcopter, and the Laubloomis benchmarks in \cite{huang2022polar}. In all case studies, we train $1$-step surrogate models that we combine into a $\horizon$-step surrogate model for trajectory prediction. For reachability analysis of the  surrogate model we use the approx-star algorithm from \cite{tran2020nnv} for the Laubloomis case study, while we use the exact-star algorithm from \cite{tran2020nnv} for the adaptive cruise controller and the quadcopter. The underlying system is described by an ordinary differential equation (ODE), potentially affected by noise, and we consider the system at discrete time points. Therefore, we let the control input and the noise be fixed over the sampling time $[\timeid \delta t, (1+\timeid)\delta t], \timeid \in [\horizon]$. The star-sets that we obtain are $n>3$ dimensional. Illustration of our results is thus demonstrated through their projection onto state components. This is important to keep in mind when assessing the tightness of the obtained star-sets.

\myipara{Adaptive Cruise Control}
We consider a $6$-dimensional system consisting of a leader and a follower vehicle that is equipped with an adaptive cruise controller. The underlying black-box model is described by the  ODE model 
\begin{equation}
\label{eq:acc}
\begin{bmatrix}\dot{x}_1\\ \dot{x}_2\\ \dot{x}_3\\ \dot{x}_4\\
\dot{x}_5\\ \dot{x}_6\end{bmatrix} =\begin{bmatrix} x_2\\ x_3\\
-2x_3-4-\mu x_2^2\\ x_5\\ x_6\\ -2x_6+2u-\mu x_4^2\end{bmatrix}+ v, \ \  \init=\left\{  \state_0\ \middle| \  \begin{bmatrix} 90\\32\\0\\10\\30\\0\end{bmatrix} \leq \state_0 \leq \begin{bmatrix} 110\\32.2\\0\\11\\30.2\\0\end{bmatrix} \right\}
\end{equation}
where $v \sim \gaussian(0, d^2)$ with standard deviation $d= \mathbf{diag}([1, 0.1,  0.05,  1,  0.1,  0.05 ])$. The friction coefficient is $\mu=10^{-4}$. The state components $x_1,x_2,x_3$ are the position, velocity, and hidden state of the lead vehicle, while the components $x_4,x_5,x_6$ are the position, velocity, and hidden state of the ego vehicle. The surrogate model is a neural network with ReLU activation functions and with  layers $[6, 32,32,32,6]$. The surrogate model is trained from a training dataset generated from Eq.~\eqref{eq:acc} where the control input $u$ is generated by the adaptive cruise controller. To perform conformal inference, we generate $L=40000$  i.i.d. $50$-step trajectories from Eq.~\eqref{eq:acc} with sampling time $\delta t =0.1 \sec$ that we collect within the test dataset $D_{\text{test}}$. 
Fig.~\ref{fig:acc_comparison} shows the projections of our probabilistic flowpipe onto its state components by setting $\varepsilon =0.01$. To visualize the tightness of the probabilistic flowpipe, we generate $100000$ additional i.i.d. trajectories from Eq.~\eqref{eq:acc} and include their projection on their state components in Fig.~\ref{fig:acc_comparison} as well. Table \ref{tbl:acc} shows the run time of our experimental results for the range $\varepsilon \in [0.01, 0.1]$.  

\begin{table*}[t]
\centering
\resizebox{\textwidth}{!}{
\begin{tabular}{ccc||ccc} 
\toprule
Failure & Conformal inference & Reachability& Failure  & Conformal inference & Reachability  \\
 Probability & Run-time & Run-time & Probability & Run-time & Run-time \\ 
\midrule \rowcolor{Gray}
$0.10$ &   2.475 $\mathrm{sec}$ &7.3779 $\mathrm{sec}$ &  $0.05$ & 2.4762 $\mathrm{sec}$ &7.5 $\mathrm{sec}$  \\
$0.09$ & 2.4754 $\mathrm{sec}$ &6.8911  $\mathrm{sec}$  &  $0.04$ & 2.4822 $\mathrm{sec}$ &7.3542 $\mathrm{sec}$  \\
\rowcolor{Gray}
$0.08$ &  2.5894 $\mathrm{sec}$ &7.2825  $\mathrm{sec}$  &  $0.03$ & 2.5316 $\mathrm{sec}$ &7.5548 $\mathrm{sec}$  \\
$0.07$ &  2.6046 $\mathrm{sec}$ &7.5339 $\mathrm{sec}$ &  $0.02$ & 2.364 $\mathrm{sec}$ &7.2504 $\mathrm{sec}$  \\
\rowcolor{Gray}
$0.06$ & 2.4674 $\mathrm{sec}$ &7.1876 $\mathrm{sec}$  & $0.01$ & 3.5858 $\mathrm{sec}$ &7.4376 $\mathrm{sec}$  \\
\bottomrule
\end{tabular}
}
\caption{Quadcopter: Computation times of our method for different user-provided failure probabilities $\varepsilon$. The reachability run-time is the overall time for reachability analysis over the surrogate model and constructing the probabilistic flowpipe with conformal inference. The data generation time for the test dataset is $273\ \sec$ and the run time for training the $\relu$ model on training dataset is  $2$ hours and $25$ minutes. }\label{tbl:Quad}
\end{table*}

\begin{table*}[t]
\centering
\resizebox{\textwidth}{!}{
\begin{tabular}{ccc||ccc} 
\toprule
Failure & Conformal inference & Reachability& Failure  & Conformal inference & Reachability  \\
 Probability & Run-time & Run-time & Probability & Run-time & Run-time \\ 
\midrule \rowcolor{Gray}
$0.10$ &   89.9305 $\mathrm{sec}$ & 93.8766 $\mathrm{sec}$ &  $0.05$ & 71.3700 $\mathrm{sec}$ & 49.7159 $\mathrm{sec}$  \\
$0.09$ &  72.0818 $\mathrm{sec}$ & 53.0224  $\mathrm{sec}$  &  $0.04$ & 67.8528
$\mathrm{sec}$ & 50.1378 $\mathrm{sec}$  \\
\rowcolor{Gray}
$0.08$ &  74.2623 $\mathrm{sec}$ & 50.0363  $\mathrm{sec}$  &  $0.03$ & 72.8819 $\mathrm{sec}$ & 50.1828 $\mathrm{sec}$  \\
$0.07$ &  71.7105 $\mathrm{sec}$ & 49.7770 $\mathrm{sec}$ &  $0.02$ & 85.9235 $\mathrm{sec}$ & 49.7792 $\mathrm{sec}$  \\
\rowcolor{Gray}
$0.06$ & 69.4149 $\mathrm{sec}$ & 50.7159 $\mathrm{sec}$  & $0.01$ & 91.9276 $\mathrm{sec}$ & 92.9743 $\mathrm{sec}$  \\
\bottomrule
\end{tabular}
}
\caption{Laubloomis: Computation times of our method for different user-provided failure probabilities $\varepsilon$. The reachability run-time is the overall time for reachability analysis over the surrogate model and constructing the probabilistic flowpipe with conformal inference.  The data generation time for the test dataset is $19$ minutes, and the run time for training the $\relu$ model on the training dataset is  $2$ hours and $30$ minutes. }\label{tbl:bench1_laub}
\end{table*}

\myipara{Quadcopter}
We consider a 12-dimensional Quadcopter model that was introduced in \cite{huang2022polar}. The ODE model for the Quadcopter is shown in Eq.~\eqref{eq:Quad12}. The additive noise $v=\left[v_1, v_2, \cdots, v_{12}\right]$ is Gaussian $v \sim \gaussian(0, d^2)$ with standard deviation $d= \mathbf{diag}([0.05\times \vec{1}_{1 \times 6},0.01\times \vec{1}_{1 \times 6}])$. The controller is a neural network controller that was presented in \cite{huang2022polar}. From this model, we generate a test dataset with  $L=40000$ data-points by sampling $50$-step trajectories with sampling time $\delta t = 0.1 \sec$.  We also train our surrogate model as a neural network with  layers $[12,20,20,20,12]$ from an additional training dataset.
\begin{figure*}
{\small 
\begin{equation}\label{eq:Quad12}
    \hspace{-3mm}\left\{ \hspace{-3mm} \begin{array}{ll}
    &\dot{x}_1 = \cos(x_8) \cos(x_9)x_4+( \sin(x_7) \sin(x_8) \cos(x_9)- \cos(x_7) \sin(x_9))x_5\\
    &\hspace{2mm}+( \cos(x_7) \sin(x_8) \cos(x_9)+ \sin(x_7) \sin(x_9))x_6+v_1\\
    &\dot{x}_2 = \cos(x_8) \sin(x_9)x_4+( \sin(x_7)  \sin(x_8)  \sin(x_9)+ \cos(x_7)  \cos(x_9)) x_5\\
    &\hspace{2mm}+( \cos(x_7)  \sin(x_8)  \sin(x_9)- \sin(x_7)  \cos(x_9)) x_6+v_2\\
    &\dot{x}_3 =  \sin(x_8) x_4- \sin(x_7)  \cos(x_8) x_5- \cos(x_7)  \cos(x_8) x_6+v_3\\
    &\dot{x}_4 = x_{12} x_5-x_{11} x_6-9.81  \sin(x_8)+v_4\\
    &\dot{x}_5 = x_{10} x_6-x_{12} x_4+9.81  \cos(x_8)  \sin(x_7)+v_5\\
    &\dot{x}_6 = x_{11} x_4-x_{10} x_5+9.81  \cos(x_8)  \cos(x_7)-9.81-u_1/1.4+v_6\\
    &\dot{x}_7 = x_{10}+( \sin(x_7) ( \sin(x_8)/ \cos(x_8))) x_{11}+( \cos(x_7) ( \sin(x_8)/ \cos(x_8))) x_{12}+v_7\\
    &\dot{x}_8 =  \cos(x_7) x_{11}- \sin(x_7) x_{12}+v_8\\
    &\dot{x}_9 = ( \sin(x_7)/ \cos(x_8)) x_{11}+( \cos(x_7)/ \cos(x_8)) x_{12}+v_9\\
    &\dot{x}_{10} = -0.9259 x_{11} x_{12}+18.5185 u_2+v_{10}\\
    &\dot{x}_{11} = 0.9259 x_{10} x_{12}+18.5185 u_3+v_{11}\\
    &\dot{x}_{12} = v_{12}
    \end{array}
    \right. 
    \begin{array}{l}
    \init= \\
    \left\{  \state_0\ \middle| \  \begin{bmatrix} -0.2\\-0.2\\-0.2\\-0.2\\-0.2\\-0.2\\0\\0\\0\\0\\0\\0\end{bmatrix} \leq \state_0 \leq \begin{bmatrix} 0.2\\0.2\\0.2\\0.2\\0.2\\0.2\\0\\0\\0\\0\\0\\0\end{bmatrix}\right\} 
    \end{array}
\end{equation}}
\end{figure*}
 The run time for our data-driven reachability analysis is shown in Table \ref{tbl:Quad} and the projections of the probabilistic flowpipes onto its state components for $\varepsilon = 0.01$ is shown in Fig.~\ref{fig:Quad}.

\begin{figure*}[t]
\hspace{-7mm}
    {\includegraphics[width=1.1\linewidth]{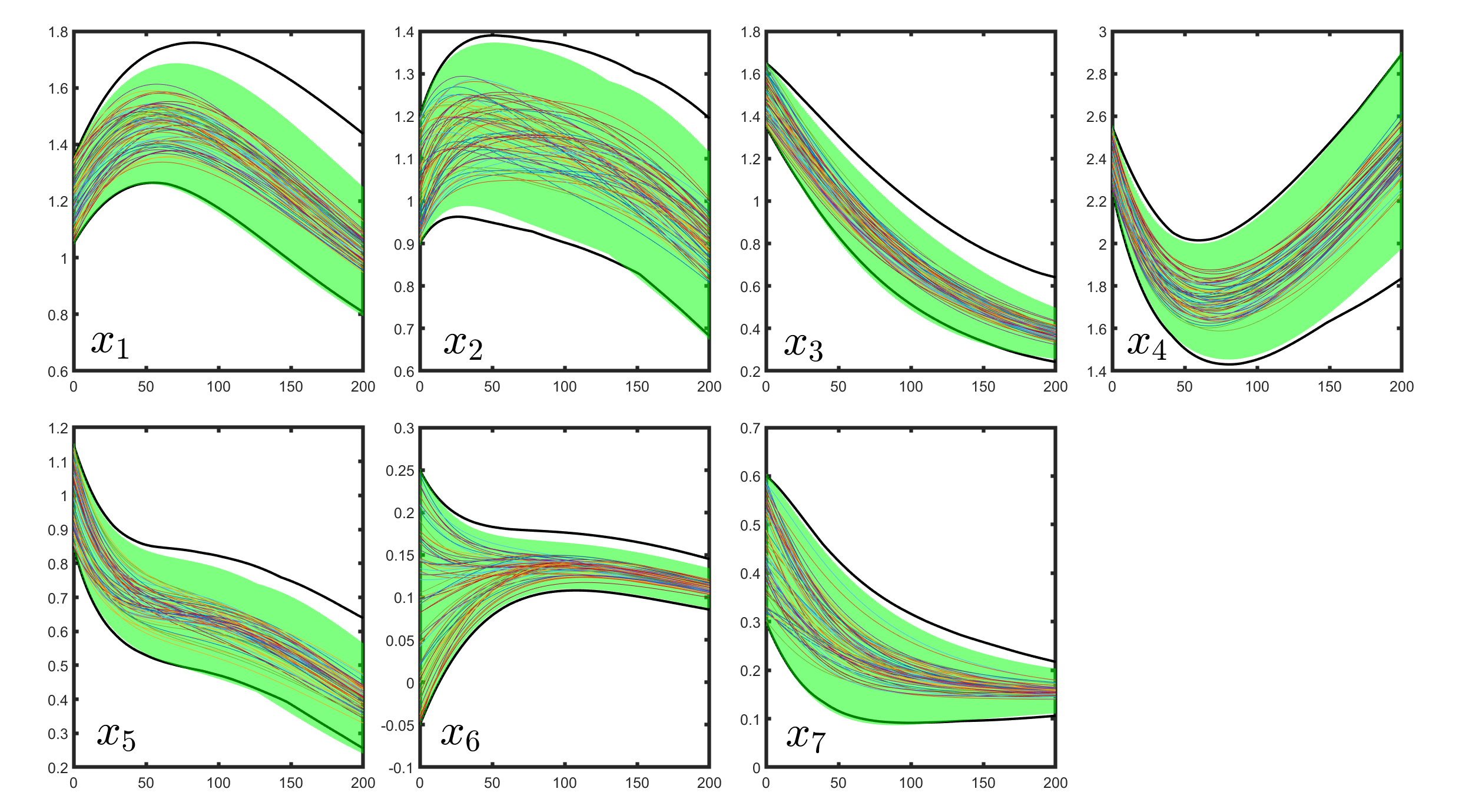}}
    {\caption{Laubloomis: Shows our data driven  reachability analysis from $200$-step dataset along with 100 random trajectories generated from $M$. We also included the reachability analysis from CORA toolbox that is based on the ideally known model depicted as green regions. The bounds in black line shows the reachability analysis with approx-star technique combined with conformal inference with prescribed failure probability $\varepsilon=0.01$.}
    \label{fig:laub}}
\end{figure*}
\myipara{Laubloomis}
Finally, we consider a  7-dimensional system which is known as Laubloomis \cite{huang2022polar}. The ODE model  for Laubloomis is shown in Eq.~\eqref{eq:laubloomis}. To provide a comparison with deterministic reachability analysis techniques, we do not add noise to the system in this case so that the system is effectively deterministic. However, note that in our approach we need to sample the initial conditions.  We thus generate a test dataset of size $L =160000$ consisting of $200$-step trajectories with sampling time $\delta t =0.01$. 

\begin{equation}
\label{eq:laubloomis}
\begin{bmatrix}\dot{x}_1\\ \dot{x}_2\\ \dot{x}_3\\ \dot{x}_4\\
\dot{x}_5\\ \dot{x}_6\\ \dot{x}_7\end{bmatrix} =\begin{bmatrix} 1.4 x_3-0.9 x_1\\ 2.5 x_5-1.5 x_2\\
0.6  x_7-0.8 x_3 x_2\\ 2.0-1.3 x_4 x_3\\ 0.7 x_1-1.0 x_4 x_5\\ 0.3 x_1-3.1 x_6 \\ 1.8 x_6-1.5  x_7 x_2
 \end{bmatrix}, \ \  \init=\left\{  \state_0\ \middle| \  \begin{bmatrix} 1.05\\ 0.9\\ 1.35\\ 2.25\\ 0.85\\ -0.05\\ 0.3 \end{bmatrix} \leq \state_0 \leq \begin{bmatrix} 1.35\\ 1.2\\ 1.65\\ 2.55\\ 1.15\\ 0.25\\ 0.6\end{bmatrix} \right\}
\end{equation}

Again, we train a surrogate model as a neural network with $\relu$ activation functions and layers $[7,20,20,20,7]$.  Since the horizon length of trajectories is long, we utilize the approx-star approach instead of the exact-star approach which is more efficient at the cost of conservatism. The simulation of the results for $\varepsilon = 0.01$ is also demonstrated in Fig.\ref{fig:laub}. The reachability run time for our data driven probabilistic approach for a range of failure probabilities $\varepsilon \in [0.01, 0.1 ]$ is shown in Table \ref{tbl:bench1_laub}. Finally, we  utilize the CORA toolbox \cite{althoff2015introduction} to compare to our results, which is shown in Fig.\ref{fig:laub}. We emphasize that  the CORA toolbox is only applicable to deterministic systems and assumes access to the model, while our approach only assumes availability of data.


\section{Conclusion }
\label{sec:conclusion}
We proposed a data-driven approach to analyze the reachability of stochastic dynamical systems. Particularly, we studied stochastic dynamical systems when no mathematical model of the system is available, and the only information available to us is data observed from the system. We showed how to compute probabilistic reach sets, so called probabilistic flowpipes, for this system from data. Probabilistic flowpipes ensure that the probability of a new trajectory not being in the flowpipe is upper bounded by a user-defined threshold. Our approach consists of first learning a surrogate model of the system from a training dataset. We then used the surrogate model to perform reachability analysis over it using existing tools for deterministic reachability analysis. To quantify the error between the surrogate model and the underlying unknown system, we finally use conformal inference on a test dataset. We illustrated our approach using three case studies.

\section*{Acknowledgments}
This work was supported by the National Science Foundation through the following grants: CAREER award (SHF-2048094), CNS-1932620,  FMitF-1837131, CCF-SHF-1932620, the Airbus Institute for Engineering Research, and funding by Toyota R\&D and Siemens Corporate Research through the USC Center for Autonomy and AI.

\bibliographystyle{IEEEtran}
\bibliography{Reference}

\addtolength{\textheight}{-2cm}   

\end{document}